# Bulk and Interface Effects Based on Rashba-Like States in Ti and Ru Nanoscale-Thick Films: Implications for Orbital-Charge Conversion in Spintronic Devices

Eduardo S. Santos*, José E. Abrão, Jefferson L. Costa, João G. S. Santos, Kacio R. Mello, Andriele S. Vieira, Tulio C. R. Rocha, Thiago J. A. Mori, Rafael O. R. Cunha, Joaquim B. S. Mendes, and Antonio Azevedo*

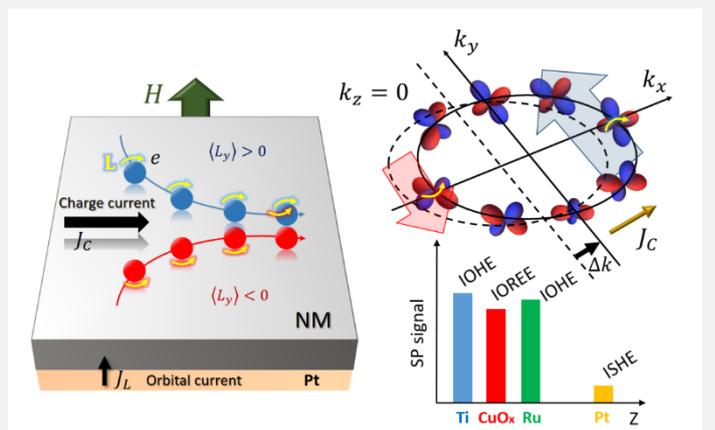

**ABSTRACT**: In this work, employing spin-pumping techniques driven by both ferromagnetic resonance (SP-FMR) and longitudinal spin Seebeck effect (LSSE) to manipulate and direct observe orbital currents, we investigated the volume conversion of spin-orbital currents into charge-current in YIG(100nm)/Pt(2nm)/NM2 structures, where NM2 represents Ti or Ru. While the YIG/Ti bilayer displayed a negligible SP-FMR signal, the YIG/Pt/Ti structure exhibited a significantly stronger signal attributed to the orbital Hall effect of Ti. Substituting the Ti layer with Ru revealed a similar phenomenon, wherein the effect is ascribed to the combined action of both spin and orbital Hall effects. Furthermore, we measured the SP-FMR signal in the YIG/Pt(2)/Ru(6)/Ti(6) and YIG/Pt(2)/Ti(6)/Ru(6) heterostructures by just altering the stack order of Ti and Ru layers, where the peak value of the spin pumping signal is larger for the first sample. To verify the influence on the oxidation of Ti and Ru films, we studied a series of thin films subjected to controlled and natural oxidation. As Cu and $CuO_x$ is a system that is already known to be highly influenced by oxidation, this metal was chosen to carry out this study. We investigated these samples using SP-FMR in YIG/Pt(2)/$CuO_x$($t_{Cu}$) and X-ray absorption spectroscopy and concluded that samples with natural oxidation of Cu exhibit more significant results than those when the $CuO_x$ is obtained by reactive sputtering. In particular, samples where the Cu layer is naturally oxidized exhibit a $Cu_2O$-rich phase. Our findings help to elucidate the mechanisms underlying the inverse orbital Hall and inverse orbital Rashba-Edelstein-like effects. These insights indeed contribute to the advancement of devices that rely on orbital-charge conversion.

## 1. INTRODUCTION

The spin Hall effect (SHE) results in a spin current perpendicular to the direction of a charge current in materials characterized by strong spin-orbit coupling (SOC).[1-3] Heavy metal films, such as Pt, Pd, W and Ta have frequently served as primary sources or detectors of spin current due to their strong SOC.[4,5] On the other hand, lighter elements like Ti and Cu, traditionally considered secondary in spintronics research, are now widely used to study spin-orbitronics phenomena at the nanoscale.[6,7] Hence, interfaces play an important role in both research and applications related to spin currents. Various techniques, including spin pumping (SP) and spin torque ferromagnetic resonance (STFMR), are employed to generate and inject spin currents across interfaces between magnetic and non-magnetic (NM) materials. SP involves the transfer of angular momentum, while STFMR uses the spin-polarized current to excite ferromagnetic resonance (FMR) in a material. SP, driven either by ferromagnetic resonance (SP-FMR) or by the application of a temperature gradient along a magnetic material (known as the spin Seebeck effect (SSE)), has emerged as one of the most employed methods to inject pure spin current into a material. While in SP-FMR the spin current injection arises from the coherent precession of the magnetization,[8,9] in SSE it occurs through the generation of a magnon current propagating along the thermal gradient.[10,11]

Over the past decade, spintronics has predominantly focused on injecting spin current into magnets for non-volatile memory applications, relying on the spin-orbit torque, which is limited to the use of materials with strong SOC. However, recent theoretical predictions and experimental discoveries have revealed the potential for orbital angular momentum (OAM) flow perpendicular to a charge current.[12-17] This phenomenon, known as the orbital Hall effect (OHE), is regarded more fundamental than the SHE, as it manifests itself independently of the SOC, distinguishing it from the SHE, which arises from the coupling between the orbital (L) and spin (S) angular momenta.[14] Consequently, there is an opportunity to integrate orbital effects with spin effects to increase the efficiency and cost-effectiveness of various spintronic devices, including non-volatile magnetic memories, magnetic sensors, nanoscale microwave sources, and other nanodevices. Such advances could contribute significantly to the evolution of the microelectronics and information technology sector.[18-20]

The OHE, intrinsic to any material with finite electron angular momentum, is an inherent property. Theoretical estimations indicate that the values for the orbital Hall conductivity ($\sigma_{OH}$) are notably larger than those of spin Hall conductivity ($\sigma_{SH}$).[13,21] As a result, both the orbital Hall effect and its inverse counterpart, the inverse orbital Hall effect (IOHE), exhibit stronger effects compared to the intrinsic spin Hall effect and its inverse, the inverse spin Hall effect (ISHE). Like SHE, the orbital counterpart comprises contributions from both bulk states (OHE) and surface states, denoted by the Orbital Rashba-Edelstein-like Effect (OREE-like), each governed by distinct underlying mechanisms. Although many theoretical studies have explored OHE and OREE, along with their inverse effects,[15-26] their generation and experimental detection have only been achieved recently.[27-39] Despite their identification and investigation, the controlled injection and accurate detection of orbital currents, specifically the flow of OAM on a nanometric scale, pose significant challenges. Spin currents and orbital currents exhibit a fundamental distinction: while spin current directly transfer torque to magnetization, orbital current does not have this ability.

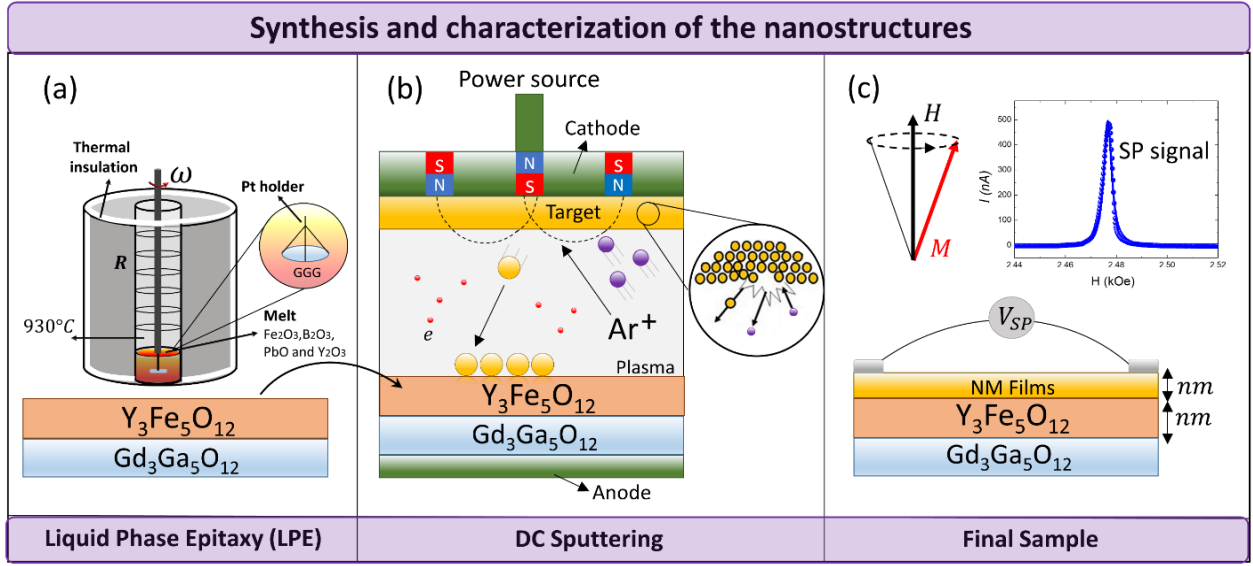

**Figure 1.** Schematic representation of the fabrication process and magnetic heterostructure measurement, illustrating (a) Liquid Phase Epitaxy (LPE) and (b) DC sputtering. High-quality YIG films are grown on GGG substrates using the LPE technique. Subsequently, the samples are cut to appropriate dimensions and transferred to the deposition chamber. DC sputtering involves injecting Ar gas into a high vacuum chamber, that is ionized and guided by a magnetic field to the target. Ar+ ions collide with the target material (e.g., Pt, Ti, Ru, or Cu), thus injecting atoms that are then deposited onto the substrate. In (c), the SP-FMR process is illustrated, wherein the YIG magnetization undergoes precession induced by an RF field. This precessing magnetization pumps a spin current that diffuses into the upper layer, subsequently converted into a DC signal via Inverse Spin Hall Effect (ISHE).

Consequently, investigation of orbital transfer torque for magnetization, especially with regard to the challenging role of natural Cu oxidation in OREE, requires further exploration.

▪ **2. RESULTS AND DISCUSSION**

In this work, we report experimental findings on the manifestation of IOHE and the Inverse Orbital Rashba-Edelstein-like Effect (IOREE) in YIG/Pt/NM1/NM2 type heterostructures, where NM1 and NM2 represent Ti or Ru. The samples used in our study start as simple bilayers until they reach multilayers, where the order and number of layers vary. In the end, we investigated the role played by a $CuO_x$ overlayer, where the *p-d* hybridization between Cu and O was investigated by X-ray absorption spectroscopy. The YIG films were grown by means of the Liquid Phase Epitaxy (LPE) technique on (111)-oriented $Gd_3Ga_5O_{12}$ (GGG) substrates, as illustrated in Figure 1(a). This technique ensures the fabrication of films characterized by exceptional quality and stability. After being grown via LPE, the YIG films were stored in a low-humidity cabinet. YIG stands out as one of the most stable magnetic insulators known to date. The subsequent metal layers were grown by means of the DC magnetron sputtering technique at room temperature under a working pressure of $2.8 \times 10^{-3}$ Torr and a base pressure of $1.7 \times 10^{-7}$ Torr or lower, which is illustrated in Figure 1(b). We began the investigation by characterizing the SP-FMR process in simple bilayers of YIG/NM1, where the NM1 layers consisted of bare films of the materials Pt, Ti, and Ru. Subsequently, we studied the interplay between spin, orbital and charge currents, in YIG/Pt(2)/NM2 heterostructures, where NM2 consisted of Ti or Ru. We further explored this phenomenon by stacking layers of Ru and Ti layers on top of YIG/Pt(2) bilayer, experimenting with different stacking order. Finally, we explore the influence of a CuO$_x$ capping layer on the previously investigated phenomenon. CuO$_x$ was generated through two distinct processes: (i) natural oxidation of sputtered Cu layers after exposure to air, or (ii) reactive sputtering in which Cu films are deposited in the presence of oxygen gas. Detailed descriptions of the SP-FMR and LSSE techniques as well as the measurement of the additional ferromagnetic damping due to the presence of the metal layer are available in the supporting information. Furthermore, other experimental details can be found in Reference (33), as identical setups were employed in this work.

- **2.1. SP-FMR and LSSE in YIG/NM1 and YIG/Pt(2)/NM2**

The inset in Figure 2(a) illustrates the experimental scheme of SP-FMR measurements. The magnetization $\vec{M}$ of the YIG layer remains fixed by the application of a dc magnetic field $\vec{H}$. When subjected to a perpendicular low magnitude $rf$ magnetic field $\vec{h}_{rf}$, $\vec{M}$ undergoes oscillations around equilibrium and reaches the resonance condition for $\vec{H} = \vec{H}_r$. Under this ferromagnetic resonance (FMR) condition, a pure spin pumping occurs across the YIG/Pt interface. This results in the generation of spin accumulation that diffuses upward through the Pt layer, causing an imbalance in the spin chemical potential, $\vec{\mu}_S$. Consequently, the strong SOC in Pt leads to an imbalance in the orbital chemical potential, $\vec{\mu}_L$. This connection between $\vec{\mu}_L$ and $\vec{\mu}_S$ is phenomenologically expressed as $\vec{\mu}_L = \delta_{LS} C \vec{\mu}_S$, where the dimensionless constant $C$ represents the strength of this relationship, and $\delta_{LS} = \pm 1$ indicates the SOC signal. This phenomenon can also be explained through the simultaneous diffusion currents of spin ($\vec{J}_S$) and orbital ($\vec{J}_L$) angular momentum, collectively represented by the intertwined spin and orbital currents $\vec{J}_{LS}$. Experimental verification demonstrates that within a NM material, the flow of these currents gives rise to a perpendicular charge current $\vec{J}_C$. This phenomenon can be attributed to the simultaneous operation of two mechanisms: ISHE (spin-related) and IOHE (orbital-related). For YIG/NM under FMR condition with NM with strong SOC, the total charge current ($\vec{J}_C^{NM}$) is expressed as the sum of ISHE induced charge current ($\vec{J}_{NM}^{ISHE}$) and the IOHE induced charge current ($\vec{J}_{NM}^{IOHE}$), $\vec{J}_C^{NM} = \vec{J}_{NM}^{ISHE} + \vec{J}_{NM}^{IOHE}$. This means that spin propagation is invariably accompanied by the flow of orbital angular momentum, particularly evident in materials that exhibit strong SOC. The spin-related contribution is given by $\vec{J}_{NM}^{ISHE} = \theta_{SH}^{NM} (\hat{\sigma}_S \times \vec{J}_S^{NM})$, where $\hat{\sigma}_S$ is the spin polarization determined by the external magnetic field $\vec{H}$, and $\theta_{SH}^{NM}$ is the spin Hall angle of the NM layer. Here, the spin current to charge current conversion occurs through SOC-induced scattering mechanisms.[2,3] Similarly, the orbital-related contribution is expressed by $\vec{J}_{NM}^{IOHE} = \theta_{OH}^{NM} (\hat{\sigma}_L \times \vec{J}_L^{NM})$, where $\hat{\sigma}_L$ represents the orbital polarization, and $\theta_{OH}^{NM}$ stands for the orbital Hall angle of the NM layer. Materials with positive $\theta_{SH}$ exhibit a spin polarization $\hat{\sigma}_S$ parallel to the orbital polarization $\hat{\sigma}_L$, i.e., $(\vec{L} \cdot \vec{S}) > 0$. On the other hand, materials with negative $\theta_{SH}$ present an antiparallel alignment between the spin polarization $\hat{\sigma}_S$ and the orbital polarization $\hat{\sigma}_L$, i.e., $(\vec{L} \cdot \vec{S}) < 0$. The mechanism of orbital-charge conversion occurs via space-momentum orbital texture, emerging from both the bulk of the materials and in systems exhibiting a broken inversion symmetry with a Rashba-type conversion.[14] The efficiency of the spin-charge and orbital-charge conversion is represented by $\theta_{SH}^{NM}$ and $\theta_{OH}^{NM}$, defined from spin Hall conductivity $\sigma_{SH}$, orbital Hall conductivity $\sigma_{OH}$, and the electrical conductivity $\sigma_e$, by $\theta_{SH}^{NM} = (2e/\hbar)\sigma_{SH}^{NM}/\sigma_e^{NM}$ and $\theta_{OH} = (2e/\hbar)\sigma_{OH}^{NM}/\sigma_e^{NM}$. Note that the resulting charge currents, generated by ISHE and IOHE, can increase or decrease based on the specific values of the orbital and spin Hall conductivities. In fact, the interplay between the spin, orbital and charge currents depends on the role played by the SOC and the spin, orbital and charge conductivities. Expressed phenomenologically, the SP-FMR signals generated in YIG/NM structures can be understood through two conversion channels: (i) spin-to-charge, attributed to ISHE, where $J_C^{ISHE} = (2e/\hbar)(\sigma_{SH}/\sigma_e)J_S$, which strongly relies on the presence of SOC, and on the ratio $\sigma_{SH}/\sigma_e$. (ii) orbital-to-charge, due to IOHE, where $J_C^{IOHE} = (2e/\hbar)(\sigma_{OH}/\sigma_e)J_L$, which is independent of SOC but depends on the existence of an orbital-texture (whether in bulk or surface) and the ratio $\sigma_{OH}/\sigma_e$. In the case of (ii), the intensity of $J_L$ can be experimentally controlled as discussed in Figure 3. The additional magnetic damping attributed to presence of a top layer of a material with strong $\sigma_{OH}$, like Ti, on YIG/Pt(2) is negligible and is discussed in Supporting Information S1.

Let's begin examining spin-charge conversion in YIG/NM1, where NM1 represents Pt, Ti, or Ru. Pt exhibits large SOC and $\sigma_{SH}^{Pt} \sim 2212\ (\hbar/e)(\Omega \cdot cm)^{-1}$, $\sigma_{OH}^{Pt} \sim 144\ (\hbar/e)(\Omega \cdot cm)^{-1}$,[21] resulting in a negligible IOHE when compared to ISHE. On the other hand, Ti has negligible SOC and $\sigma_{SH}^{Ti} \sim -17\ (\hbar/e)(\Omega \cdot cm)^{-1}$, $\sigma_{OH}^{Ti} \sim 4304\ (\hbar/e)(\Omega \cdot cm)^{-1}$,[21] resulting in a negligible ISHE signal. The comparison between ISHE signals generated by Pt and Ti can be seen in Figures 2(a) and 2(b). The ratio between the ISHE signals generated in YIG/Pt(4) and YIG/Ti(4) is given by $(I_{YIG/Pt}^{ISHE}/I_{YIG/Ti}^{ISHE}) \cdot (rf_{Ti}/rf_{Pt}) \approx -2 \times 10^3$. Here we considered the different $rf$ powers used to excite the SP-FMR signals. We also explored the conversion of spin and orbital currents to charge current in YIG/Ru, which deserves a more detailed explanation. It has been predicted,[21] that Ru has $\sigma_{SH}^{Ru} \sim 135\ (\hbar/e)(\Omega \cdot cm)^{-1}$, $\sigma_{OH}^{Ru} \sim 5545\ (\hbar/e)(\Omega \cdot cm)^{-1}$,[21] along with a SOC strength for Ru estimated to be one-third of that of Pt.[40] In this scenario, the SP-FMR signal observed in YIG/Ru can be explained by two channels. In the first channel, the spin current injected into Ru is converted into charge current. However, owing to the weak spin conductivity of Ru the resulting ISHE signal is negligible. In a second channel, the substantial SOC in Ru leads to induction of orbital states by spin states in Ru, where $\vec{J}_L^{Ru} \propto \delta_{LS} C \vec{J}_S^{Ru}$. Consequently, the resulting SP-FMR signal in Ru can be expressed by $\vec{J}_{Ru}^{SP-FMR} = \vec{J}_{Ru}^{ISHE} + \vec{J}_{Ru}^{IOHE}$ (where $\vec{J}_{Ru}^{ISHE} \ll \vec{J}_{Ru}^{IOHE}$), given that the SOC scattering of Ru shares the same polarity as in Pt. Figure 2(c) shows the SP-FMR signal for YIG/Ru(4), measured for $\phi = 0°$ (blue symbols) and $\phi = 180°$ (red symbols), with an $rf$ power of 15 mW. Comparatively, the SP-FMR signal in YIG/Ru(4) is lower than the signal generated in YIG/Pt(4) and exhibits the same polarity. Figure 2(d) shows a comparison between the SP-FMR signals for YIG/Ti(4) (red symbols), YIG/Ru(4) (black symbols) and YIG/Pt(4) (green symbols) for an $rf$ power of 15 mW at $\phi = 0°$. Note that the SP-FMR signal of YIG/Pt(4) is much larger than the SP-FMR signal of YIG/Ti(4). Therefore, SP-FMR measurements in YIG/NM1 provide important information about the SOC. From our experimental results, we can state that

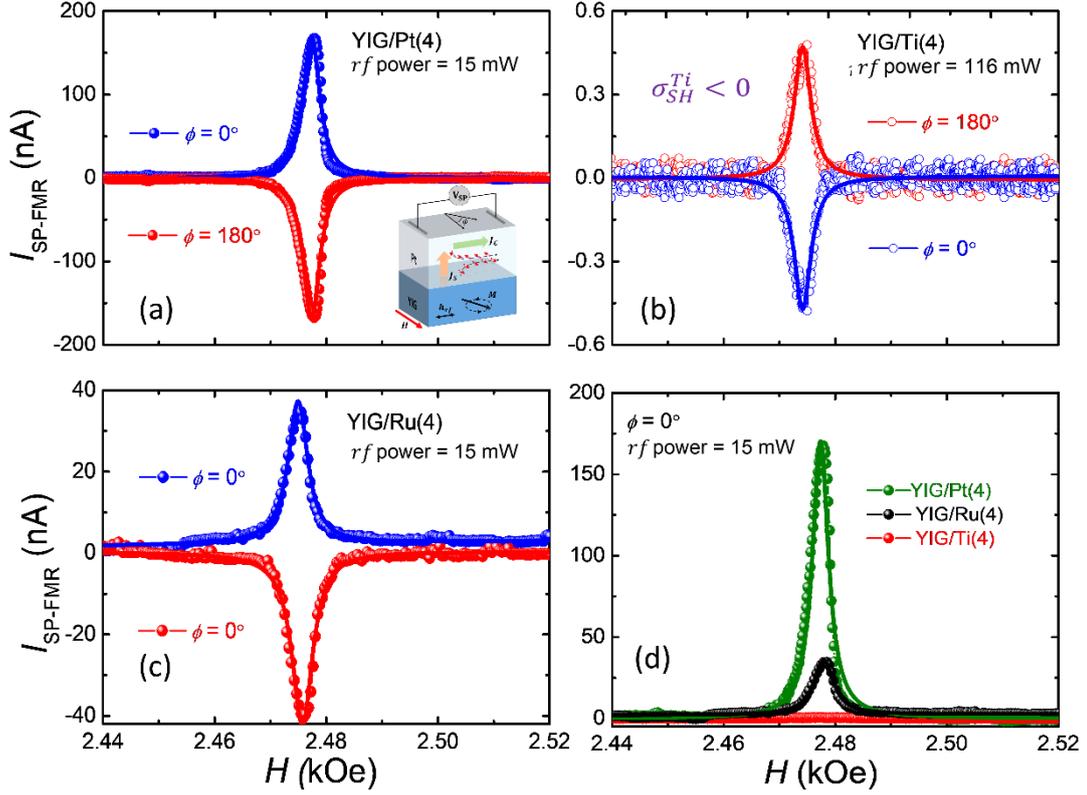

**Figure 2.** SP-FMR signals: (a) charge current measured in YIG/Pt(4), using an $rf$ power of 15 mW. The inset illustrates the SP-FMR technique where a pure spin current ($\vec{J}_S$) is injected into Pt. The SOC of Pt creates a transversal charge current that is majority given by $\vec{J}_C = \theta_{SH}^{Pt}(\hat{\sigma}_S \times \vec{J}_S)$, since $\sigma_{SH}^{Pt} \gg \sigma_{OH}^{Pt}$. (b) Charge current measured in YIG/Ti(4), employing an $rf$ power of 116 mW. Considering that the $rf$ excitation power are different, $(I_{YIG/Pt}^{ISHE}/I_{YIG/Ti}^{ISHE}) \cdot (rf_{Ti}/rf_{Pt}) \approx -2 \times 10^3$. Observe that $\sigma_{SH}^{Ti}$ and $\sigma_{SH}^{Pt}$ have opposite polarity, where positive ISHE current for Pt and the negative ISHE current for Ti, is measured for $\phi = 0°$. (c) SP-FMR charge current measured in YIG/Ru(4). In (d) we present a comparison of SP-FMR signals for YIG/Ti(4) (red symbols), YIG/Ru(4) (black symbols) and YIG/Pt(2) (green symbols), using an $rf$ power of 15 mW.

---

Pt has a strong SOC, Ru has an intermediate SOC, and Ti has a negligible SOC. Furthermore, a subtle asymmetry is observed in the signal corresponding to the YIG resonance. This asymmetry suggests that excessive RF power was employed to drive the FMR condition, leading to the onset of nonlinear effects in the YIG. Given the minor nature of the asymmetry, we used the Lorentzian curve equation, $I_{SP-FMR} = (I_s \Delta H^2)/((H - H_r)^2 + \Delta H^2)$ to adjust the experimental data. Here, $I_s$ is the signal amplitude, $H_r$ is the resonance field, and $\Delta H$ is the FMR linewidth.

Our experimental scheme has the advantage of injecting orbital current into the NM2 layer in a controlled way. This process occurs when the intertwined spin and orbital current, generated within Pt reaches the Pt/NM2 interface, thus injecting both currents into the NM2 layer. Note that the 2 nm Pt layer thickness is thin enough to allow the transit of the intertwined spin and orbital currents to the Pt/NM interface. Consequently, we observed a remarkable result in the measured SP-FMR and LSSE signals when we added a NM2 layer on top of the YIG/Pt(2). In Figure 3 several noticeable features emerge: (i) The orbital current injected into NM2 is converted into a transversal charge current via the IOHE, analogous to the ISHE, expressed as $\vec{J}_C^{NM} = \theta_{OH}^{NM}(\hat{\sigma}_L \times \vec{J}_L^{NM})$, where $\theta_{OH}^{NM} = (2e/\hbar)(\sigma_{OH}^{NM}/\sigma_{NM})$ and $\vec{J}_L^{NM}$ is the orbital

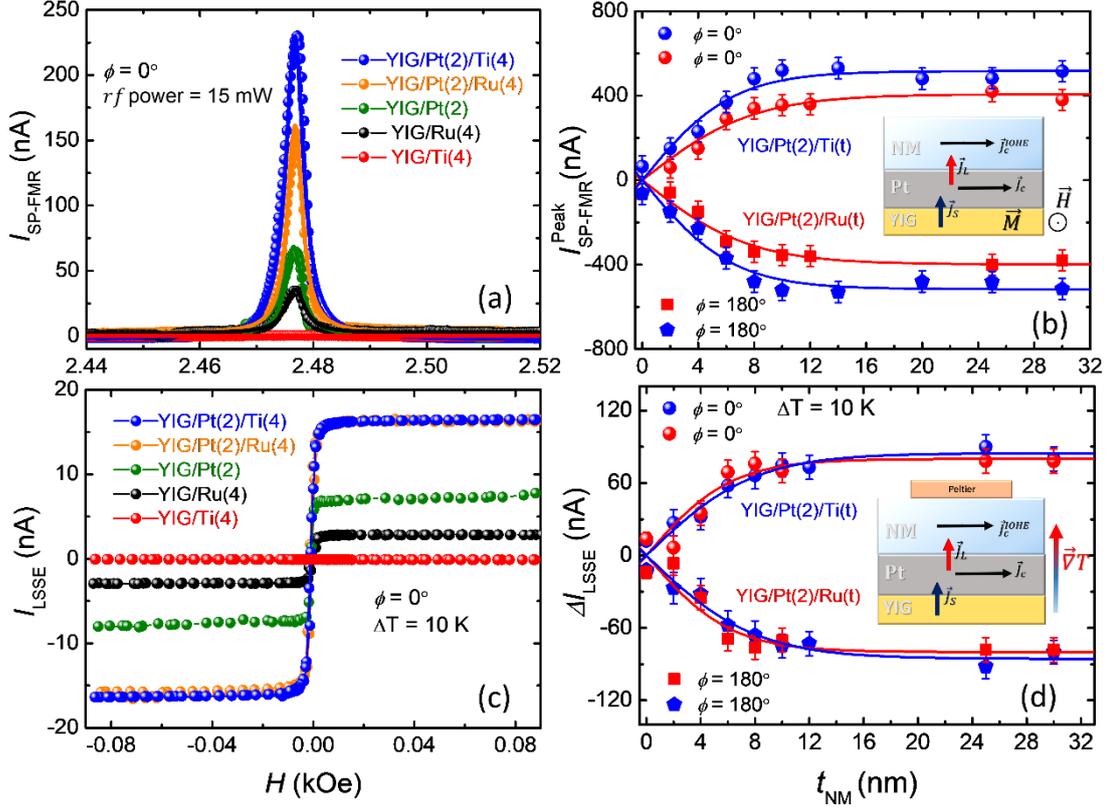

**Figure 3.** (a) Comparison of SP-FMR signals among YIG/Ti(4) (red symbols), YIG/Ru(4) (black symbols), YIG/Pt(2) (green symbols), YIG/Pt(2)/Ru(4) (orange symbols) and YIG/Pt(2)/Ti(4) (blue symbols). Lorentzian curve fits demonstrate enhanced SP-FMR signals attributed to orbital-charge conversion by IOHE in Ti and IOHE in Ru. (b) The peak values SP-FMR measurements for YIG/Pt(2)/Ti($t_{Ti}$) (blue symbols) and YIG/Pt(2)/Ru($t_{Ru}$) (red symbols) for $\phi = 0°$ and $\phi = 180°$, using 15 mW $rf$ power. The inset represents the heterostructure used to inject orbital current in NM2 from the FMR condition. (c) LSSE curves for YIG/Ti(4) (red symbols), YIG/Ru(4) (black symbols), YIG/Pt(2) (green symbols), YIG/Pt(2)/Ru(4) (orange symbols) and YIG/Pt(2)/Ti(4) (blue symbols). Notably, YIG/Ti(2) exhibited no measurable LSSE signal. (d) $\Delta I_{LSSE}$ as a function of the thickness of the Ti or Ru layer for $\Delta T = 10K$. Blue symbols are the data for NM2=Ti and the red symbols for NM2=Ru. The theoretical adjustment is given by $\Delta I_{LSSE} = A\tanh(t_{NM}/2\lambda_L)$. $\Delta I_{LSSE}$ is defined as the difference between the values measured at the magnetization saturation condition. The inset represents the heterostructure to inject orbital current in NM2 from the thermal flow of magnons from YIG.

---

current injected across the Pt/Ti interface. (ii) The resultant SP-FMR signal exhibited a gain of 3.9-fold (for YIG/Pt(2)/Ti(4) (blue symbols)) and 2.5-fold (for YIG/Pt(2)/Ru(4) (orange symbols) in comparison to that of YIG/Pt(2) (green symbols), as shown by curves in Figure 3(a). (iii) Both the polarization of orbital and spin currents align in the same direction, agreeing with the prediction of ref.[21] It is important to note that the orbital current polarization is established by spin current polarization, which in turn is dictated by the direction of the YIG magnetization oriented parallel to the external applied field. Actually, the blue signal and the orange signal of Figure 3(a) represent the effective charge current $\vec{J}_C^{eff}$ in YIG/Pt(2)/NM2(4), which is given by $\vec{J}_C^{eff} = \theta_{SH}^{Pt}(\hat{\sigma}_S \times \vec{J}_S^{Pt}) + \theta_{OH}^{NM}(\hat{\sigma}_L \times \vec{J}_L^{NM})$. This effectively illustrates the combined effect of ISHE of Pt and IOHE of NM2.

We then explored the conversion of spin and orbital currents into charge current by varying the thickness of the Ti and Ru layers from 0 to 30 nm. In Figure 3 (b), the maximum values of SP-FMR signals are plotted as a function of the NM layer thickness for YIG/Pt(2)/Ti($t_{Ti}$) (blue symbols) and YIG/Pt(2)/Ru($t_{Ru}$) (red symbols). Both signals clearly saturate beyond 12 nm thickness. Comparing the SP-FMR signal values at the saturation region for YIG/Pt(2)/Ti(30) and YIG/Pt(2)/Ru(30) with YIG/Pt(2), significant gains were observed: $I^{SP-FMR}_{YIG/Pt(2)/Ti(30)} / I^{ISHE}_{YIG/Pt(2)} \approx 7.6$ and $I^{SP-FMR}_{YIG/Pt(2)/Ru(30)} / I^{ISHE}_{YIG/Pt(2)} \approx 6$. Given that $\sigma_{OH} > 0$ for Ti and Ru, the signals generated by IOHE add to the ISHE signal generated in YIG/Pt(2). Furthermore, note that our experimental results show that $\hat{\sigma}_L$ and $\hat{\sigma}_S$ are parallel in Pt (SOC > 0), which leads to gains in the SP-FMR or LSSE signals due to IOHE in Ti or Ru, in $\phi = 0°$, and at $\phi = 180°$, according to the equation: $\vec{J}^{eff}_C = \theta^{Pt}_{SH}(\hat{\sigma}_S \times \vec{J}^{Pt}_S) + \theta^{NM}_{OH}(\hat{\sigma}_L \times \vec{J}^{NM}_L)$, where $\theta^{Pt}_{SH} > 0$ and $\theta^{NM}_{OH} > 0$, confirming the theoretical results available in the literature.[21]

The experimental data of Figures 3(b) and 3(d) were fitted using the equation $I^{SP-FMR}_{NM} = A\tanh(t_{NM}/2\lambda_L)$,[39] where $A$ is a constant, $t_{NM}$ is the thickness of the NM2 material and $\lambda_L$ represents the orbital diffusion length. From the adjustment to the experimental data, we found $\lambda^{Ti}_L = (3.2 \pm 0.4)$ nm and $\lambda^{Ru}_L = (3.8 \pm 0.6)$ nm, which are larger than the spin diffusion length in Pt, where $\lambda^{Pt}_S \sim 1.6$ nm, or smaller.[33] According to,[33,41] spin pumping (by SP-FMR or LSSE) in YIG/Pt samples creates a flow of spin angular momentum in Pt, which, due to the strong SOC, is accompanied by a collinear orbital angular momentum flow that can be injected in an adjacent NM2 layer. Although the difference in diffusion lengths between Ti and Ru is small, the significant SOC of Ru compared to Ti could elucidate the larger $\lambda_L$ of Ru.[39,41] This may seem counterintuitive, as SOC typically leads to the dissipation of angular momentum. However, as indicated in,[41] the dissipation of S and L is explained by the parameters $\lambda_S$ and $\lambda_L$, respectively, while the additional phenomenological parameter $\lambda_{LS}$ describes the non-dissipative exchange between orbital and spin angular momentum. Thus, $\lambda_{LS}$ effectively expands the spatial range of orbital and spin accumulation. Therefore, despite the relatively weaker SOC of Ru compared to Pt, it still outperform that of Ti.[40] Thus an increased orbital diffusion length is anticipated in samples with Ru films. However, both models, the phenomenological theory[39,41] and the orbital diffusion theory,[42] need to be confirmed with a first-principles theory that explains orbital relaxation.

To validate the reliability of measurements performed by SP-FMR, we examined the conversion of spin and orbital currents to charge current using the longitudinal SSE (LSSE) technique. Figure 3(c) shows the results of the LSSE measurements performed on the following set of samples: YIG/Ti(4), YIG/Ru(4), YIG/Pt(2), YIG/Pt(2)/Ru(4), and YIG/Pt(2)/Ti(4). Similar to the results obtained by the SP-FMR technique, enhancements in LSSE signals were observed when adding Ti or Ru over YIG/Pt(2), by applying temperature difference of 10 K between top and bottom sample surface. Figure 3(d) presents the results of LSSE measurements as a function of the thickness of the Ti and Ru films. The inset of Figure 2(d) schematically shows the LSSE technique, where a temperature gradient $\vec{\nabla}T$ is applied vertically. The Peltier module in contact with the top surface gives the high temperature, such that the spin current $\vec{J}_S$ is injected upward through the YIG/Pt interface, where $\vec{J}_S \rightarrow \vec{J}_C$ due to ISHE in Pt. The spin current results from difference in magnon population along the YIG film.[11] In the case of thermal SP, the underlying physics is like the SP-FMR phenomenon. Despite the SP-FMR values for YIG/Pt(2)/Ti($t_{Ti}$) consistently surpassing those for YIG/Pt(2)/Ru($t_{Ru}$), as displayed in Figure 3(b), the SP currents obtained by SSE showed equivalent values, regardless of whether the Ti or Ru layer covered the Pt layer. Moreover, it was observed that the LSSE signal for YIG/Ru(4) bilayer is relatively smaller compared to the LSSE observed in YIG/Pt(2). It is noteworthy that despite the measurable SP-FMR signal exhibited by the YIG/Ti(4) sample (Figure 2(b)), its LSSE signal was negligible and indistinguishable from the background noise (red symbols in Figure 3(c)). The difference between the results regarding the conversion of spin and orbital currents to charge current obtained by SP-FMR and SSE techniques arises from a fundamental contrast between the two processes. The SP-FMR spin injection relies on the coherent rotation of magnetization at the YIG/Pt interface. On the other hand, spin injection by applying a temperature gradient is intrinsically incoherent due to its thermal nature. As a result, SP-FMR is expected to be more efficient in spin injection compared to the LSSE process.

Intriguingly, our SP-FMR measurements indicate that Ti exhibits a more significant contribution than Ru, which seems to contradict the theoretical results.[21] It is important to acknowledge that our samples deviate from ideal single crystalline films, thereby preventing a direct qualitative comparison with theoretical predictions. Also, it is worth mention that Ru films have larger electrical conductivity ($\sigma_e$) than Ti films. From the measurement of sheet resistance ($R_s$) we found that $\sigma_e^{Ru}/\sigma_e^{Ti} \sim 5.5$. According to the definition $\theta_{OH} = 2e/\hbar(\sigma_{OH}/\sigma_e)$, this implies that Ru should exhibits a lower orbital-charge conversion efficiency compared to Ti. Lastly, it is crucial to emphasize that our work constitutes an experimental study, providing valuable information to inform and guide future refinements in theoretical investigations.

### ▪ 2.2. Propagation of orbital currents along multilayer

To investigate the propagation of orbital current, as well as its conversion to charge current, throughout multilayer systems, we performed SP-FMR and LSSE measurements on heterostructures with varying number of layers and stack orders. Figure 4(a) shows SP-FMR measurements conducted on the YIG/Pt(2)/Ru(6)/Ti(6) (sample $\alpha$) and YIG/Pt(2)/Ti(6)/Ru(6) (sample $\beta$) heterostructures, showing the impact of changing the stack order of Ti and Ru layers. The peak value of the SP-FMR signal is around 930 nA and 680 nA for samples $\alpha$ and $\beta$, respectively. The enhancement in the SP-FMR signal of the sample $\alpha$ relative to YIG/Pt(2) is about 14 times, while for sample $\beta$ it is about 11 times. Figure 4(b) shows the LSSE signals for sample $\alpha$ and sample $\beta$, for a temperature difference between the bottom and top surfaces of 10 K. The obtained LSSE signals exhibited enhancements, compared to YIG/Pt(2), of approximately 5.3 times and 6.4 times for samples $\alpha$ and sample $\beta$, respectively.

Our results reveal that the signals of YIG/Pt(2)/Ti are slightly larger than YIG/Pt(2)/Ru. This small difference must be directly linked to the orbital Hall angle $\theta_{OH}$. Despite Ru exhibiting significant SOC in comparison to Ti, the efficiency of orbital-to-charge conversion within Ti films is higher. The heterostructures in the form of YIG/Pt(2)/NM1/NM2 stacks added more complexity to the study. In the case of the YIG/Pt(2)/Ru(6)/Ti(6) sample, the observed increase in signals can be attributed to the higher $\theta_{OH}$ exhibited by Ti compared to Ru. Since the Pt layer facilitates the injection of orbital and spin currents

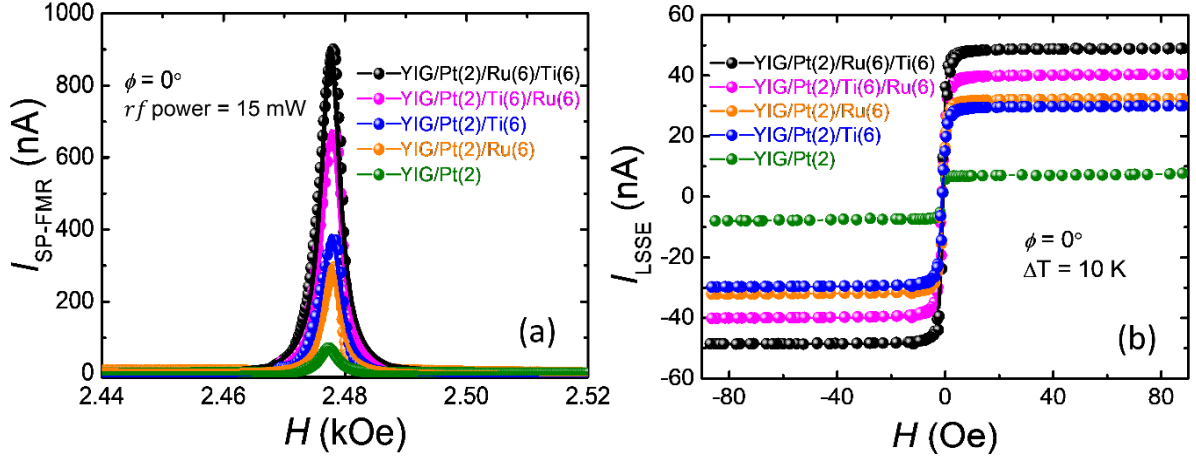

**Figure 4**. (a-b) SP-FMR and LSSE signals, respectively, for YIG/Pt(2)/Ru(6)/Ti(6) (black symbols), YIG/Pt(2)/Ti(6)/Ru(6) (pink symbols), YIG/Pt(2)/Ti(6) (blue symbols) and YIG/Pt(2)/Ru(6) (orange symbols) heterostructures compared with signals for YIG/Pt(2) (green symbols). By just changing the stack order of Ti and Ru layers, we obtained an expressive gain in the SP-FMR and the LSSE signals.

---

to the adjacent layer, the conversion to charge current is more pronounced within the Ti layer than in Ru. On the other hand, in the YIG/Pt(2)/Ti(6)/Ru(6) sample, the enhancement is less substantial, mainly due to the weaker spin-orbit coupling (SOC) of Ti, resulting in most of the injection orbital arising only from the Pt(2) layer. It is crucial to note that these explanations are highly qualitative and require theoretical elucidation from the first principles. Our experimental findings are expected to stimulate future theoretical investigations.

The findings in this section highlight that orbital currents exhibit the ability to propagate over long distances compared to spin currents. This is supported by the observation that despite employing multiple layers of diverse materials, intended to convert orbital current into charge current, noticeable enhancement in the SP-FMR and LSSE signals were consistently observed.

- **Characterization of oxidized layers and IOREE-like measurements**

It is well known that Ti and Ru form highly stable oxides. Given the absence of a capping layer in the YIG/Pt/NM2 samples, it is conceivable that the formation of oxides on Ti and Ru could also impact orbital transport. To gain deeper insights into the influence of metal layer oxidation, we conducted a comprehensive study involving a range of thin films subjected to controlled and natural oxidation processes. Taking advantage of the well-studied Pt/CuO$_x$ interface as a model system to investigate the Rashba-type orbital effect,[27,33] we prepared a series of samples like those used in previous sections, including a naturally oxidized capping layer of copper (Cu). To investigate the optimum thickness of the CuO$_x$ capping layer, we prepared a series of samples of YIG/Pt(2)/Cu($t_{Cu}$) for $0 \leq t_{Cu} \leq 8$ nm. After two days of natural oxidation, we measured the SP-FMR signal generated by each sample. The average peak values, calculated between the data of $\phi = 0°$ and $\phi = 180°$, are shown in Figure 5(a), as a function of the Cu layer thickness. As shown, the sample YIG/Pt(2)/CuO$_x$(3) exhibited the largest SP-FMR signal with a gain $I^{SP-FMR}_{YIG/Pt(2)/CuOx(3)}/I^{SP-FMR}_{YIG/Pt(2)} \approx 5.3$, in agreement with previously published results.[33] It is important to mention that the YIG/Pt(2)/Cu($t_{Cu}$) samples with $t_{Cu}$ between 1 and 3 nm have a higher $I^{Peak}_{SP-FMR}$ value than the samples with $t_{Cu} > 4$ nm. This slight decrease can indeed be attributed to the oxygen depletion diffusing into the Cu layer and reaching the Pt/Cu interface, a phenomenon that certainly needs further investigation. Regarding the role played by the natural oxidation of Ti and Ru in the results shown in Figure 3, it is evident that both SP-FMR and LSSE signals are predominantly influenced by bulk effects. If these signals were dependent on the surface oxidation, we would expect to observe in thin regime of Ti and Ru, a different dependence than that shown in Figure 3. Additionally, Figure 5(b) shows the sheet resistance ($R_S$) measurement of the sample Si/Cu(3) left to naturally oxidize. These measurements were performed automatically at half-hour intervals. We observe that naturally oxidized Cu films retain metallic characteristics, maintaining finite electrical continuity. This is evident even in thinner Cu layers with thickness of 3nm, that ensures complete oxygen penetration throughout the Cu thickness. We observed a sharp increase in electrical resistance from 100 to 120 ohms within the initial 5 hours and then slowly increases to 140 Ohms over the subsequent 48 hours.

To evaluate the depth of the oxidation process in the Cu layer, we investigated the cross-section of the GGG/Cu/Pt samples using transmission electron microscopy (TEM) and atomic resolution energy-dispersive x-ray spectroscopy (EDS). Figure 5(c) shows the typical cross-section EDS mapping analysis of Cu films after naturally oxidation for two days. The result shows the atomic distributions of the atoms: gadolinium (Gd) and gallium (Ga) (present in the GGG substrate), copper (Cu), oxygen (O), platinum (Pt) and gold (Au). Note that Pt and Au (on top) come from protective layers grown after the Cu oxidation process and that were necessary for the procedure of cross-section lamella preparation using Focused Ion Beam (FIB) technology. The atomic percentage of each layer was confirmed by EDS line profile as shown in Figure 5(d). Scanning was done along the yellow arrow in Figure 5(c) and shows the atomic distribution percentages of the elements Cu and O in typical samples investigated. The TEM and EDS results confirmed the existence of an oxidation layer on the surface of the Cu films, as these reveal a substantial presence of O in the surface layer of Cu, which can be up to ~10 nm wide. The Figure 5(d) guarantees the presence of oxygen even for thicknesses greater than 3 nm, which can be directly associated with Figure 5(a), that is, for all Cu thicknesses we used there was a considerable presence of oxygen throughout the Cu thickness.

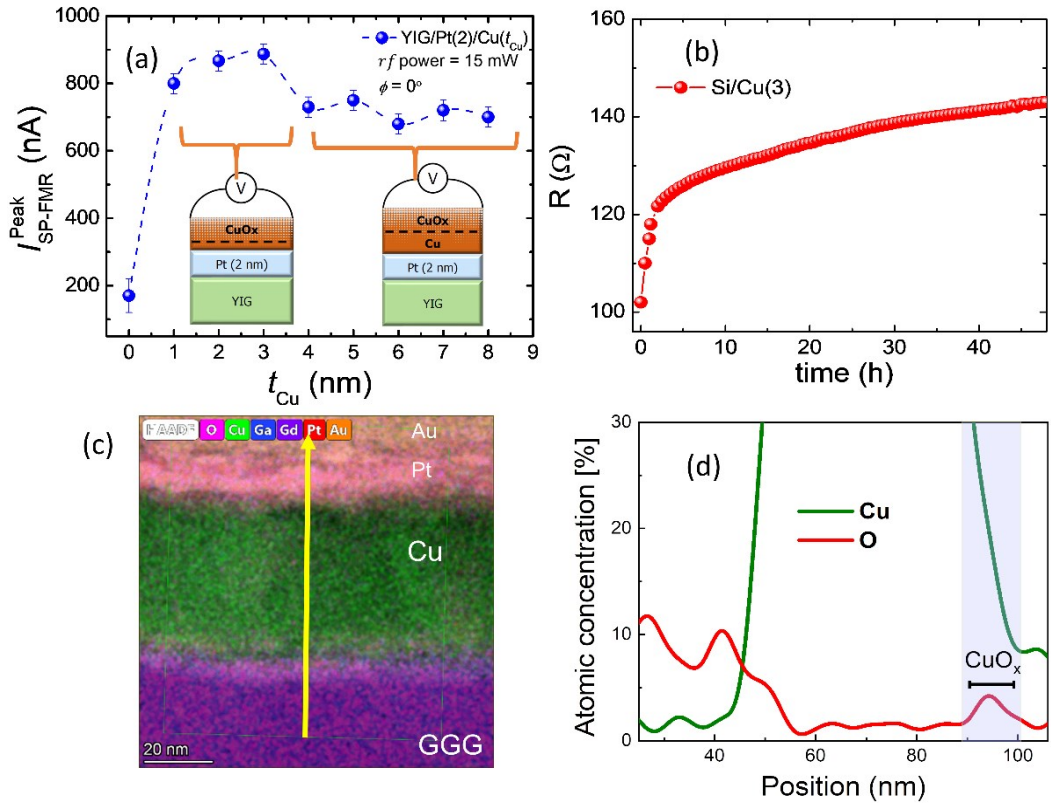

**Figure 5.** (a) Average values of the SP-FMR peak signal for YIG/Pt(2)/Cu($t_{Cu}$), as function of the Cu layer thickness, measured after naturally oxidation for two days. The dashed line is just a guide for the eye. (b) shows the sheet resistance ($R_S$) of a Si/Cu(3) as a function of the oxidation time. (c) Typical cross-section EDS mapping analysis of Cu films after naturally oxidation for two days. The result shows the atomic distributions of the atoms: Ga and Gd (present in the GGG substrate), Cu, O, Pt and Au. Note that Pt and Au (on top) come from protective layers grown after the Cu oxidation process and that were necessary for the procedure of cross-section lamella preparation using FIB. (d) Line scan taken along the yellow arrow in (c). Atomic distribution of the elements Cu and O is illustrated by their corresponding atomic percentages, displaying a substantial presence of oxygen, with an approximate width of up to ~10 nm.

---

The combined effect of the IOHE and IOREE-like will be presented in Figure 7 further, where we explored both SP-FMR and LSSE in a series of samples capped with a CuO$_x$(3) layer.

- **X-ray absorption spectroscopy**

Involving the transition of a 2p core electron to unoccupied 3d states above the Fermi level, the X-ray absorption L-edge of transition metals is representative of the electronic structure around the valence levels. Thus, we carried out X-ray absorption spectroscopy (XAS) measurements, at the Cu L$_3$-edge (around 930 eV) of Cu-based model samples, to gain more insights into the chemical nature of the oxide formed in oxidized thin films. The measurements were carried out at the inelastic scattering and photoelectron spectroscopy (IPE) beamline of the Sirius light source at the Brazilian Synchrotron Light Laboratory,[43] using both total electron yield (TEY) and fluorescence yield (FY) acquisition modes. The TEY signal is surface sensitive, probing just a few nanometers from the surface. The weight of the contribution from each atomic layer to the TEY spectrum exponentially decreases with depth. On the other hand, the FY mode provides a spectrum that is averaged along the full depth of the thin film since the photon-in/photon-out signal is bulk sensitive.

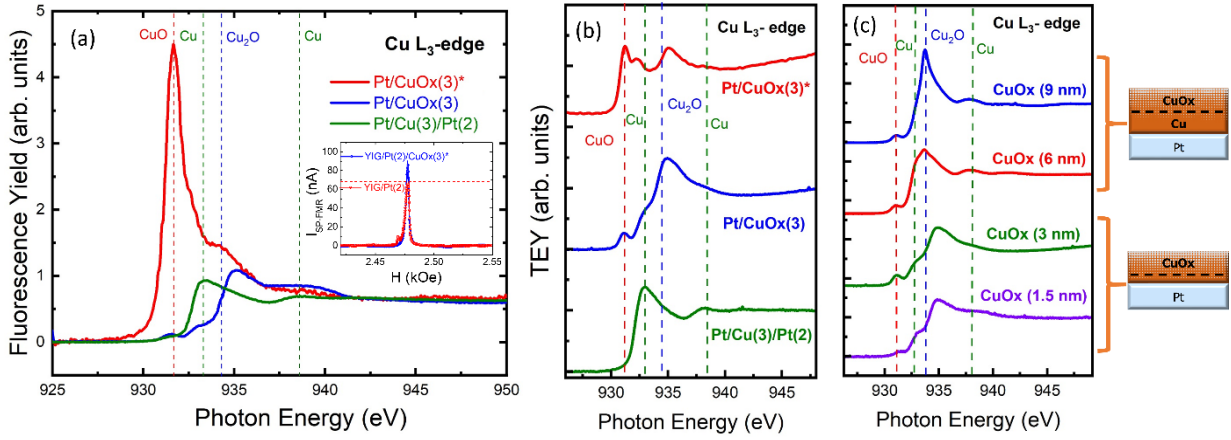

**Figure 6.** (a) XAS spectra acquired by FY for three samples with the following structures: Pt/CuO$_x$(3)* (red), Pt/CuO$_x$(3) (blue) and Pt/Cu(3)/Pt(2) (green). The inset shows a comparison between the SP-FMR signals of YIG/Pt(2)/CuO$_x$(3)* (blue) and YIG/Pt(2) (red) for $\phi = 0°$ and rf power = 15 mW. (b) XAS spectra acquired by TEY mode. (c) TEY spectra of naturally oxidized Cu thin films of thicknesses ranging from 1.5 nm to 9.0 nm.

Figure 6(a) exhibits the XAS spectra acquired by FY for three samples with the following structures: Pt/CuO$_x$(3)* (red), Pt/CuO$_x$(3) (blue) and Pt/Cu(3)/Pt(2) (green). While the copper oxide layer of the first sample was deposited by applying oxygen with a 5.3% flow ratio into argon gas during the sputtering, the one of the second sample was naturally oxidized after the deposition of metallic Cu. In its turn, the third sample is a reference metallic stack with a Pt capping layer deposited on Cu to prevent oxidation. The spectrum of this reference sample is consistent with that found in the literature for the L$_3$-edge of metallic copper,[44] presenting two characteristic features around 933 and 938 eV (green dashed line in Figure 6(a)). It is well-known that CuO (Cu$^{2+}$) shows a strong peak placed around 2 eV bellow the Cu$^0$ metal edge (red dashed line), and Cu$_2$O (Cu$^{1+}$) exhibits a peak around 1 eV above it (blue dashed line). Therefore, it is clear in Figure 5(a) that the sample grown by reactive sputtering is highly oxidized, the spectrum being dominated by CuO. On the other hand, the naturally oxidized Cu film spectrum seems to exhibit a combination of partially oxidized and metallic copper, appearing features from both CuO, Cu$_2$O, and metallic Cu. The inset of Figure 6(a) shows a comparison between the SP-FMR signals of YIG/Pt(2)/CuO$_x$(3)* and YIG/Pt(2). Note that there is no significant gain in the signal when adding CuO$_x$(3)*. According to ref.,[26] the OREE arises from the $pd$ hybridization between Cu and O valence orbitals that leads to a momentum space OAM texture in partially oxidized Cu interfaces.

The XAS spectra acquired by TEY mode are shown in Figure 6(b). In the case of the sample grown by reactive sputtering (red), the CuO main peak appears much weaker than in the FY spectrum. This discrepancy is due to radiation damage, as the X-ray incidence under high vacuum pressure induces a reduction in the oxidation state. We observe that this effect takes place in the same timescale of the spectrum acquisition and is more remarkable in the TEY spectrum. Thus, the oxide reduction diffuses from the surface to the bulk. In contrast, although the spectrum of the naturally oxidized Cu film (blue) is dominated by $Cu_2O$, it presents a slightly higher CuO peak in the TEY spectrum. This suggests that the oxidation in ambient atmosphere leads to a gradual decrease of the oxygen content from the surface to the bulk. We further investigated the depth profile of the natural oxide through the measurement of samples with different thicknesses.

Figure 6(c) presents the TEY spectra of naturally oxidized Cu thin films of thicknesses ranging from 1.5 to 9.0 nm. Although the spectra are dominated by $Cu_2O$ and present a small peak addressed to CuO, the features arising from metallic Cu (green dashed line) are stronger for thicker samples. This result corroborates the presence of an oxidation gradient along the Cu film. While the samples with thicknesses of 1.5 and 3.0 nm appear to exhibit a partially oxidized state of Cu throughout their thickness, the samples with 6.0 and 9.0 nm seem to maintain a more metallic Cu state in the region closer to the Pt/Cu interface, as we represent in the inset of Figure 6(c). The presence of metallic Cu layers may be responsible for the reduction in SP-FMR signals for $t_{Cu} > 3$ nm due to deviations of the orbital current along the metallic Cu layer. It is important to mention that a deeper investigation of the electronic structure of partially oxidized Cu thin films is needed to unequivocally determine the mechanism behind the enhancement of the orbital transport in these samples. Nevertheless, it is noteworthy that the proposed picture is in accordance with the SP-FMR results shown in Figure 5(a).

- **SP-FMR and LSSE in YIG/Pt(2)/NM2/CuO$_x$(3)**

We decided to investigate the SP-FMR and LSSE signals in YIG/Pt(2)/NM2(4)/CuO$_x$(3) heterostructures to analyze the behavior of orbital currents along the NM2 layer and the orbital-charge conversion. Figure 7(a) and Figure 7(b), exhibit a comparison of SP-FMR signals between YIG/Pt(2), YIG/Pt(2)/NM2, and YIG/Pt(2)/NM2/CuO$_x$(3). When comparing the SP-FMR signals obtained from YIG/Pt(2)/Ti(4)/CuO$_x$(3) and YIG/Pt(2)/Ru(4)/CuO$_x$(3) with those from YIG/Pt(2), significantly higher gains of approximately 10.0-fold and 14.0-fold, are respectively observed. Notably, the introduction of a CuO$_x$ capping layer results in an enhancement of the SP-FMR signals, compared to the samples lacking CuO$_x$ coverage, with ratios of $I^{SP-FMR}_{YIG/Pt(2)/Ti(4)/CuOx(3)}/I^{SP-FMR}_{YIG/Pt(2)/Ti(4)} \approx 3.7$ and $I^{SP-FMR}_{YIG/Pt(2)/Ru(4)/CuOx(3)}/I^{SP-FMR}_{YIG/Pt(2)/Ru(4)} \approx 6$. The substantial enhancement observed in Ru(4)/CuO$_x$(3) can be attributed to the larger SOC of Ru in contrast to the negligible SOC of Ti. The remarkable gain observed after the introduction of the CuO$_x$ layer arises from the residual orbital current that reaches the NM2/CuO$_x$ interface, where it undergoes conversion into an additional charge current by the IOREE-like, generating an extra gain in the SP-FMR signal. In this case the effective charge current is given by $\vec{J}^{eff}_C = \theta^{Pt}_{SH}(\hat{\sigma}_S \times \vec{J}^{Pt}_S) + \theta^{NM}_{OH}(\hat{\sigma}_L \times \vec{J}^{NM}_L) + \vec{J}^{IOREE}_{CuOx}$, with $\vec{J}^{IOREE}_{CuOx} = \lambda_{IOREE}(\hat{z} \times \delta\vec{L})$, where $\lambda_{IOREE}$ is the efficiency orbital-charge conversion by Rashba-like states, and $\delta\vec{L}$ represents the non-equilibrium orbital density caused by orbital injection in NM/CuO$_x$ interface. The same effect of increasing the resultant charge current by combining of IOHE and IOREE-like, was observed by measuring the LSSE signal generated by the thermal-driven spin pumping effect. Figure 7(c) and Figure 7(d) show the LSSE signals after capping the YIG/Pt(2)/NM2 heterostructures with a CuO$_x$(3) layer. For a temperature difference between the bottom and top surfaces of 10 K, the LSSE signals obtained from YIG/Pt(2)/Ti(4)/CuO$_x$(3) and YIG/Pt(2)/Ru(4)/CuO$_x$(3) with those from YIG/Pt(2), showed gains of approximately 5.5-fold and 10.0-fold, respectively.

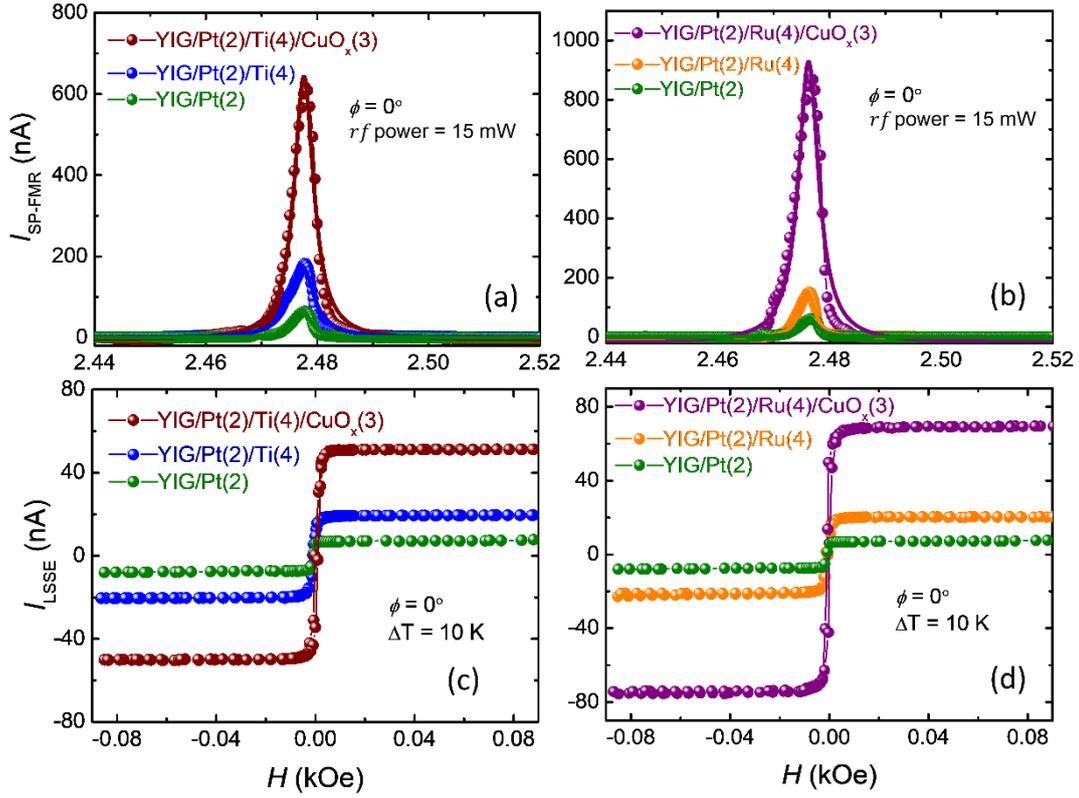

**Figure 7.** Panels (a) and (b) show the SP-FMR signals of YIG/Pt(2)/NM2 heterostructures measured with and without the CuO$_x$ capping layer, respectively. Panels (c) and (d) depict the LSSE signals of the same heterostructures, measured with and without the CuO$_x$ capping layer, respectively. The observed enhancements of the SP-FMR and LSSE signals, in the same heterostructures, using the two different techniques, are attributed to the combined effects of the ISHE, IOHE and IOREE-like, as discussed in the text.

According to the Figures 5 and 6, oxygen easily penetrates 3nm of Cu. Consequently, the chemical structure of YIG/Pt(2)/Ti and YIG/Pt(2)/Ti/CuO$_x$(3) samples should not show any disparity, since oxygen oxygen reaches the Ti layers equally in both samples. Furthermore, the monotonic variation of the SP-FMR signals as a function of the Ti layer thickness (see Figures 3 (b) and 3(d)) reveals that the bulk effects predominantly influence Ti behavior.

### ▪ 3. CONCLUSIONS

In conclusion, we investigated the effects of IOHE and IOREE on YIG/Pt/NM1/NM2 and YIG/Pt/NM1/CuO$_x$ heterostructures, where NM1 and NM2 represent nanometer thick films of Ti or Ru. Our study shows the relevance of bulk IOHE in Ti and Ru films, wherein the spin current injected through the YIG/Pt interface undergoes transformation into an intertwined spin and orbital current. The degree of entanglement observed depends on the spin-orbit coupling of the material. Interestingly, as the thickness of both the Ti and Ru layers increased, we observed that the IOHE signals reached saturation beyond 12 nm thickness. Employing a phenomenological analysis, we determined that the orbital diffusion lengths for Ru and Ti vary slightly. Furthermore, our experimental characterizations of naturally and reactively oxidized Cu layers revealed complex structures characterized by different oxidation states of Cu. Remarkably, naturally oxidized Cu predominantly exhibited the $Cu_2O$ state, while reactively oxidized Cu was dominated by the CuO state. We verified that there is no significant amplification of the signals due to IOREE when using reactively oxidized Cu; however, substantial gains were observed with natural oxidized Cu. A notable finding of our study was the remarkable enhancement of SP-FMR and LSSE signals by more than a 10-fold factor upon addition of a naturally oxidized layer of Cu(3) on top of YIG/Pt(2)/NM2(4) heterostructures. This highlights the fundamental role of IORRE in converting orbital currents into charge currents at the nanometric scale. This work certainly contributes to the advancement of materials physics and chemistry in the field of orbitronics by elucidating the intricate interactions among spin, charge and orbital degrees of freedom. These insights not only promise improvements in the efficiency of existing spintronic devices, but also pave the way for the development of new nanoelectronic devices that take advantage of orbital current flow.

## ▪ ASSOCIATED CONTENT

**Data Availability Statement**

The data that support the findings of this study are available from the corresponding author upon reasonable request.

**Supporting Information**

Section S1, analysis of magnetic damping in YIG, YIG/Pt and YIG/Pt/Ti heterostructures; Section S2, short summary of the FMR process that we used for the SP-FMR measurements (PDF)

## ▪ AUTHOR INFORMATION

**Corresponding Authors**

*__Eduardo S. Santos__ − Departamento de Física, Universidade Federal de Pernambuco, 50670-901, Recife, Pernambuco, Brazil. orcid.org/0000-0002-1413-2376; Email: edu201088@hotmail.com

*__Antonio Azevedo__ − Departamento de Física, Universidade Federal de Pernambuco, 50670-901, Recife, Pernambuco, Brazil. orcid.org/0000-0001-8572-9877; Email: antonio.azevedo@ufpe.br

**Authors**

**José E. Abrão** - Departamento de Física, Universidade Federal de Pernambuco, 50670-901, Recife, Pernambuco, Brazil. orcid.org/0000-0002-7463-1476; Email: elias_abrao@hotmail.com

**Jefferson L. Costa** - Departamento de Física, Universidade Federal de Pernambuco, 50670-901, Recife, Pernambuco, Brazil. orcid.org/0009-0006-9118-7190; Email: jefferson.limacosta@ufpe.br

**João G. S. Santos** - Departamento de Física, Universidade Federal de Pernambuco, 50670-901, Recife, Pernambuco, Brazil. orcid.org/0000-0001-8654-3564; Email: joao.gustavos@ufpe.br

**Kacio R. Mello** - Departamento de Física, Universidade Federal de Pernambuco, 50670-901, Recife, Pernambuco, Brazil. orcid.org/0000-0003-1234-1964; Email: kacioreinaldo@gmail.com

**Andriele S. Vieira** - Departamento de Física, Universidade Federal de Viçosa, 36570-900, Viçosa, Minas Gerais, Brazil. orcid.org/0000-0002-7473-4934; Email: andriele.vieira@ufv.br

**Tulio C. R. Rocha** - Laboratório Nacional de Luz Síncrotron (LNLS), Centro Nacional de Pesquisa em Energia e Materiais (CNPEM), 13083-970, Campinas, São Paulo, Brazil. orcid.org/0000-0001-5770-8366; Email: tulio.rocha@lnls.br

**Thiago J. A. Mori** - Laboratório Nacional de Luz Síncrotron (LNLS), Centro Nacional de Pesquisa em Energia e Materiais (CNPEM), 13083-970, Campinas, São Paulo, Brazil. orcid.org/0000-0001-5340-3282; Email: thiago.mori@lnls.br

**Rafael O. R. Cunha** - Departamento de Física, Universidade Federal de Viçosa, 36570-900, Viçosa, Minas Gerais, Brazil. orcid.org/0000-0002-6039-3892; Email: rafael.cunha@ufv.br

**Joaquim B. S. Mendes** - Departamento de Física, Universidade Federal de Viçosa, 36570-900, Viçosa, Minas Gerais, Brazil. orcid.org/0000-0001-9381-0448; Email: joaquim.mendes@ufv.br

- **ACKNOWLEDGMENTS**

This research is supported by Conselho Nacional de Desenvolvimento Científico e Tecnológico (CNPq), Coordenação de Aperfeiçoamento de Pessoal de Nível Superior (CAPES), Financiadora de Estudos e Projetos (FINEP), Fundação de Amparo à Ciência e Tecnologia do Estado de Pernambuco (FACEPE), Universidade Federal de Pernambuco, Multiuser Laboratory Facilities of DF-UFPE, Fundação de Amparo à Pesquisa do Estado de Minas Gerais (FAPEMIG) - Rede de Pesquisa em Materiais 2D and Rede de Nanomagnetismo, and INCT of Spintronics and Advanced Magnetic Nanostructures (INCT-SpinNanoMag), CNPq 406836/2022-1. This research used the facilities of the Brazilian Synchroton Light Laboratory (LNLS) and Brazilian Nanotechnology National Laboratory (LNNano), part of the Brazilian Centre for Research in Energy and Materials (CNPEM), a private nonprofit organization under the supervision of the Brazilian Ministry for Science, Technology, and Innovations (MCTI). We thank the IPE beamline of the LNLS/CNPEM for the synchrotron beamtimes (proposal 20232791), and the LNNano/CNPEM for advanced infrastructure and technical support during sample preparation by Focused Ion Beam (FIB) and measurements by Transmission Electron Microscopy (TEM). The TEM staff is acknowledged for their assistance during the experiments (Proposals No. 20210467 and 20230795, TEM-Titan facility).